\documentclass[11pt]{article}
\usepackage{amssymb}
\usepackage{citesort}
\begin{document}
\begin{flushright}
                LPT Orsay \\
                August 2003  \\
                hep-th/0308131
\end{flushright}
\bigskip
\begin{center}
{\LARGE\bf\sf Polyakov conjecture for hyperbolic singularities}
\end{center}
\bigskip
\begin{center}
    {\large\bf\sf
    Leszek Hadasz}\footnote{e-mail: hadasz@th.u-psud.fr} \\
\vskip 1mm
    Laboratoire de Physique Th\`{e}orique, B{\^a}t. 210,\\
    Universit{\'e} Paris-Sud,  91405 Orsay, France \\
        {\em and }\\
    M. Smoluchowski Institute of Physics, \\
    Jagiellonian University,
    Reymonta 4, 30-059 Krak\'ow, Poland \\
\vskip 5mm
    {\large\bf\sf
    Zbigniew Jask\'{o}lski}\footnote{e-mail: Z.Jaskolski@if.uz.zgora.pl
}\\
\vskip 1mm
    Physics Institute\\
    University of Zielona G\'{o}ra\\
    ul. Szafrana 4a,
    65-069 Zielona G\'{o}ra, Poland
\end{center}
\vskip .5cm
\begin{abstract}
We propose the form of the Liouville action satisfying Polyakov
conjecture on the accessory parameters for the hyperbolic
singularities on the Riemann sphere.
\end{abstract}
\newpage
\vskip 1cm
\section{Introduction}

Let us consider the Fuchsian equation
\begin{equation}
\label{Fuchs}
\partial^2\psi(z) + \frac{1}{2}T(z)\psi(z) = 0 \ ,
\end{equation}
where
\begin{equation}
\label{T:zz}
T(z) = \sum\limits_{j=1}^{n}\left[\frac{\Delta_j}{(z-z_j)^2} +
\frac{c_j}{z-z_j}\right].
\end{equation}
In the context of the Liouville field theory $T(z)$ plays the role
of the $zz$-component of the energy--momentum tensor and the
real positive numbers
 $\Delta_j$ are conformal weights. The complex numbers $c_j$
are  called accessory parameters.
The requirement that $T(z)$ is regular at the infinity
implies the relations
\begin{equation}
\label{restrictions}
\sum\limits_{j=1}^{n}c_j = 0 \ , \hskip 1cm
\sum\limits_{j=1}^{n}z_jc_j = -
\sum\limits_{j=1}^{n}\Delta_j  \ , \hskip 1cm
\sum\limits_{j=1}^{n}z_j^2c_j = -
2\sum\limits_{j=1}^{n}z_j \Delta_j \ .
\end{equation}
The Polyakov conjecture concerns the following
version of the Riemann-Hilbert problem \cite{Cantini:2001wr,Cantini:2001im,Cantini:2002jw}.
For a given set of positive weights $\{\Delta_j\}_{j=1}^n$
one has to adjust
the accessory parameters in such a way that the Fuchsian equation
(\ref{Fuchs})
admits a fundamental system of solutions
with  $SU(1,1)$ monodromies around all singularities.

The interest to this problem comes from its close relation
to the  Liouville equation on the punctured Riemann sphere
$X=(\,\mathbb{C}\cup \{\infty \}\,)\setminus \{z_1,\dots ,z_n\}$.
If $\chi_1(z), \chi_2(z)$ are linearly independent solutions of
(\ref{Fuchs})
then the function $\varphi(z,\bar z)$ determined by the relation
\begin{equation}
\label{phi}
e^{\varphi(z,\bar z)} \; = \;
{4\,|w'|^2 \over (1 - |w|^2 )^2} \ ,
\hskip 5mm
w(z) \,=\, {\chi_1(z)\over\chi_2(z)} \ ,
\end{equation}
satisfies the Liouville equation
\begin{equation}
\label{EOM}
\partial\bar\partial\varphi = \frac{1}{2} {\rm e}^\varphi
\end{equation}
for all that $z$ for which $w(z)$ is well defined. It is convenient to use
normalized solutions with Wronskian satisfying:
\begin{equation}
\label{Wronskian}
\partial\chi_1(z)\,\chi_2(z)
-\chi_1(z)\,\partial\chi_2(z)
\; = \; 1 \ ,
\end{equation}
so that the relation (\ref{phi}) can be written in a simple form
\begin{equation}
\label{phi:pol2}
e^{-{\varphi(z,\bar z)\over 2}} \; = \;\pm\;{1\over 2}
\left(
\overline{\chi_2(z)}\chi_2(z)
-\overline{\chi_1(z)}\chi_1(z)
\right).
\end{equation}
Note that $\varphi(z,\bar z)$ is real by construction.
If in addition  $\chi_1(z), \chi_2(z)$ satisfy the $SU(1,1)$ monodromy
condition then $\varphi(z,\bar z)$ is single-valued.

Under some restrictions on conformal weights  the relation above
can be made more precise.
The case
 of all parabolic singularities  was  analyzed  by Poincar\'e
in the context of the uniformization
problem \cite{Poincare}.
He showed that
the Liouville equation (\ref{EOM}) has a unique real-valued regular on $X$ solution
with the following behavior at the punctures:
$$
\varphi(z,\bar z) = \left\{
\begin{array}{lll}
-2\log | z- z_j | - 2 \log| \log | z-z_j || + O(1) & {\rm as } & z\to z_j\ , \\
-4\log | z- z_j | + O(1) & {\rm as } & z\to \infty\ .
\end{array}
\right.
$$
This solution defines a metric $ds^2 = {\rm e}^{\varphi} |dz|^2$ which is complete on $X$.
It has constant negative curvature $-1$  and  parabolic singularities at each $z_j$.
The energy--momentum tensor of the solution $\varphi,$
\begin{equation}
\label{T:parabolic}
T_\varphi(z)
\; = \;
-\frac12(\partial\varphi)^2 + \partial^2\varphi
 \; = \;
-2{\rm e}^{\frac{\varphi}{2}}\partial^2 {\rm e}^{-\frac{\varphi}{2}},
\end{equation}
is a holomorphic function on $X$ of the form (\ref{T:zz}) with the conformal weights
$$
\Delta_j = \frac{1}{2},
\hskip 5mm
j=1,\dots,n.
$$
It follows from (\ref{T:parabolic})
that there exists a pair of solutions $\chi_1,\chi_2$
to the Fuchsian equation
(\ref{Fuchs})  related to $\varphi$ by (\ref{phi})
\cite{Cantini:2001wr,Cantini:2001im,Cantini:2002jw,ZoTa,Takhtajan:2001uj}.
Since $\varphi$ is real and single-valued this solves the
$SU(1,1)$ Riemann-Hilbert problem.

The  existence and the uniqueness of the solution to the Liouville
equation with the elliptic singularities,
$$
\varphi(z,\bar z) = \left\{
\begin{array}{lll}
-2\left(1-{\theta_i\over 2 \pi} \right)\log | z- z_j |  + O(1) & {\rm as } & z\to z_j \ , \\
-4\log | z- z_j | + O(1) & {\rm as } & z\to \infty \ ,
\end{array}
\right.
$$
was proved by Picard \cite{Picard1,Picard2} (see also \cite{Troyanov} for the modern proof).
The solution can be interpreted as the conformal factor of the complete, hyperbolic metric on $X$
with the conical singularities
of the opening angles $0<\theta_j<2\pi$
at the punctures $z_j$. The energy--momentum tensor
takes the form (\ref{T:zz}) with
$$
\Delta_j = \frac{1}{2} - \frac{1}{2}\left({\theta_j\over 2 \pi} \right)^2,
\hskip 5mm
j=1,\dots,n \ .
$$
As in the parabolic case one can show
that there exists a solution to the corresponding $SU(1,1)$ monodromy problem
\cite{Cantini:2001wr,Cantini:2001im,Cantini:2002jw,Bilal:Gervais,ZoTa,Takhtajan:2001uj}.

The Polyakov conjecture states that the (properly defined and normalized) Liouville
action functional $S[\phi]$ evaluated on the classical solution $\varphi(z,\bar z)$
is a generating function for the accessory parameters of
the monodromy problem described above :
\begin{equation}
\label{generating}
c_j = - \frac{\partial S[\varphi]}{\partial z_j} \ .
\end{equation}
This formula was derived within path integral approach to the quantum Liouville theory as
the quasi-classical limit of the conformal Ward identity \cite{Polyakov82,Takhtajan:yk,Takhtajan:1994vt}.
In the case of the parabolic singularities on $n$-punctured Riemann sphere a rigorous proof based
on the theory of quasiconformal mappings was given by Zograf and Takhtajan \cite{ZoTa}.
It was also pointed out that the same technique applies for the elliptic singularities of finite order.
Note that only in these two cases the monodromy group of the Fuchsian equation
is (up to conjugation in $SL(2,\mathbb{C})$) a discrete subgroup in $SU(1,1)$,
and the map $w(z)$ defined by (\ref{phi}) solves the uniformization problem \cite{ZoTa}.

An alternative method, working both in the case of parabolic and general elliptic
singularities, was recently developed
by Cantini, Menotti and Seminara \cite{Cantini:2001wr,Cantini:2001im,Cantini:2002jw}.
Yet another proof, based on a direct calculation of the regularized Liouville action
for parabolic and general elliptic singularities,
was proposed by Takhtajan and Zograf in \cite{Takhtajan:2001uj}.

The aim of this Letter is to
find the action which satisfies (\ref{generating}) for the singularities of the
hyperbolic type. The $SU(1,1)$ monodromy problem
is well posed in this case, but whether it has a solution for arbitrary conformal weights $\Delta_j>\frac12$
and arbitrary locations of punctures $z_j$ is up to our knowledge an open problem.
In the present Letter we assume that such solution exists.

 As one can expect from the case of two
punctures  \cite{sei} the corresponding solution to the Liouville equation determined by the
relation (\ref{phi}) develops concentric line
singularities around each puncture. We assume that these line singularities do not intersect
and that they are the only singularities of the classical solution.

The singular behavior  around  punctures can be described in terms of local conformal maps.
This allows
for the construction of an appropriately regularized Liouville
action functional. One can then apply the method of Takhtajan and
Zograf \cite{Takhtajan:2001uj} to prove the Polyakov conjecture.
 The problem of the existence  and the uniqueness of the solution to the
Fuchsian equation  with the properties stated above goes beyond the scope of the
present letter.
Let us only mention that in the  case of three hyperbolic singularities
an explicit solution in terms of the hypergeometric functions exists \cite{Bilal:Gervais,HadJas4}.
There is also a simple geometrical construction yielding a large class
of solutions with an arbitrary number of
hyperbolic singularities \cite{HadJas4}.

The choice of the Fuchsian equation (\ref{Fuchs}) as a starting point for the construction
of the Liouville field theory is a convenient way to impose the crucial constraint on  admissible
classical solutions --- the holomorphic form (\ref{T:zz}) of their energy--momentum tensor $T(z)$.
We hope that the singular hyperbolic solutions and the corresponding Liouville action
will be helpful in understanding
the factorization problem in the geometrical approach to the Liouville theory developed
by Takhtajan  \cite{Takhtajan:yk,Takhtajan:1994vt,Takhtajan:1993vt,Takhtajan:zi}.
This was actually our main motivation for the present paper.

Finally let us  note that the hyperbolic solutions provide multi black hole
solutions of the 3-dim gravity \cite{Banados:wn,Brill:1995jv,Welling:1997fw}.

\section{Hyperbolic singularities}

Let us assume that $\{ \chi_1,\chi_2 \} $ is a normalized
solution to the $SU(1,1)$ monodromy
problem for hyperbolic weights
$$
\Delta_j = {1+\lambda_j^2\over 2}\ , \hskip 5mm \lambda_j \in {\mathbb R} \ .
$$
Then the fundamental system defined by
$$
\left(\begin{array}{c}
\psi_1\\
\psi_2
\end{array}\right)=
\sqrt{\frac{i}{2}}
\left(\begin{array}{cr}
1 & 1\\ i & -i
\end{array}\right)
\left(\begin{array}{c}
\chi_1\\
\chi_2
\end{array}\right)
$$
is also normalized and has
$SL(2,\mathbb{R})$ monodromy around all punctures.
In terms of $\{\psi_1,\psi_2\}$ the formula (\ref{phi:pol2}) reads
\begin{equation}
\label{phi:pol3}
e^{-{\varphi(z,\bar z)\over 2}} \; = \;\pm\;\frac{i}{2}
\left(
\psi_1(z)\overline{\psi_2(z)}
-\overline{\psi_1(z)}\psi_2(z)
\right).
\end{equation}
The advantage of using solutions with  $SL(2,\mathbb{R})$ monodromy
is that any  hyperbolic  element $M\in SL(2,\mathbb{R})$ (${\rm tr}(M)>2$)
is $SL(2,\mathbb{R})$-conjugate to a diagonal matrix.
Thus for each singularity $z_j$ there exists
an element $B_j\in SL(2,\mathbb{R})$ such that  the system
$$
\left(\begin{array}{c}
\psi^j_+\\
\psi^j_-
\end{array}\right)=
B_j
\left(\begin{array}{c}
\psi_1\\
\psi_2
\end{array}\right)
$$
has a diagonal monodromy at $z_j$.
It follows that $\psi^j_\pm$ have the canonical form
\begin{eqnarray}
\label{psis}
\psi^j_\pm(z) & = & {{\rm e}^{ \pm i{\vartheta_j\over 2}} \over \sqrt{i\lambda_j} }
(z-z_j)^{\frac{1\pm  i\lambda_j}{2}} u^j_\pm (z) \ ,
\end{eqnarray}
where $ \vartheta_j \in \mathbb{R}$,
and $ u^j_\pm(z)$ are analytic functions
\begin{eqnarray*}
\label{psi:expansion}
u^j_\pm (z)
&  = & \sum\limits_{l=0}^\infty\;
u^j_{\pm,\,l }\ (z-z_j)^{\,l} \ ,
\hskip 5mm
u^j_{\pm,\,0}\; = \; 1 \ ,
\end{eqnarray*}
on the disc $D_j=\{z\in X: |z-z_j| < \min\limits_{i,\ i\neq j} |z_i-z_j|\}$.
Expanding the energy--momentum tensor
$$
T(z) \; = \; \sum\limits_{j=1}^{n}\left[\frac{\Delta_j}{(z-z_j)^2} +
\frac{c_j}{z-z_j}\right]  \; = \; \sum\limits_{k=0}^\infty\; t_k^j\ (z-z_j)^{k-2},
$$
one gets
\begin{equation}
\label{T:coef}
t^j_0 \; = \; {\Delta_j}\ , \hskip 5mm
t^j_1 \; = \; c_j\ ,\hskip 5mm
t^j_k \; = \; \sum\limits_{i,\ i\neq j}
\left[\frac{(k-1)\Delta_i}{(z_i-z_j)^k} -
\frac{c_i}{(z_i-z_j)^{k-1}}\right]
\hskip 5mm {\rm for}\; k \geqslant 2\ .
\end{equation}
The Fuchsian equation (\ref{Fuchs}) then implies
\begin{equation}
\label{u:coef}
u^j_{\pm,\,l }\; = \; -\frac{1}{2 l (1 \pm i\lambda_j)}
\sum\limits_{k=1}^l\; t^j_k u^j_{\pm,\,l -k}
\hskip 5mm {\rm for}\;\; l \geqslant 1 \ .
\end{equation}
It is a well known property of the Schwarz derivative
\[
\{f(z),z\} \; \equiv \; \frac{f^{(3)}(z)}{f'(z)} -
\frac32\left(\frac{f''(z)}{f'(z)}\right)^2
\]
and the Fuchsian equation (\ref{Fuchs}) that the ratio
\[
A_j(z) = \frac{\psi^j_+(z)}{\psi^j_-(z)}
\]
satisfies the relation
\begin{equation}
\label{Schwarz}
T(z) \; = \; \{A_j(z),z\} \ .
\end{equation}
For each hyperbolic singularity we  define
\begin{equation}
\label{rho}
\rho_j(z) = \left(A_j(z)\right)^{\frac{1}{i\lambda_j}} .
\end{equation}
It follows from (\ref{psis},\ref{T:coef},\ref{u:coef}) that $\rho_j(z)$
is an analytic  function on $D_j$ and:
\begin{equation}
\label{rho:explicit}
\rho_j(z) \; = \; {\rm e}^{\vartheta_j\over \lambda_j}
\left[z-z_j + \frac{c_j}{2\Delta_j}(z-z_j)^2 +
{\cal O}\left((z-z_j)^3\right)\right].
\end{equation}
Using (\ref{Schwarz}) and the properties of the Schwarz
derivative one gets
\begin{equation}
\label{trans}
T(z) \; = \; \left\{\left(\rho_j(z)\right)^{i\lambda_j},z\right\}
\; = \; \left(\frac{d\rho_j(z)}{dz}\right)^2 \widetilde T^j\left(\rho_j(z)\right)
 + \{\rho_j(z),z\} \ ,
\end{equation}
where
\begin{equation}
\label{T:0}
 \widetilde T^j(\rho) \; = \; \left\{\rho^{i\lambda_j},\rho\right\}
\; = \;
\frac{\Delta_j}{\rho^2} \ .
\end{equation}
Let us consider the Fuchsian equation
$$
\partial^2\psi(\rho) + \frac{1}{2}\widetilde T^j(\rho)\psi(\rho ) = 0
$$
on the complex $\rho$ plane and a normalized fundamental system of solutions
with the diagonal  monodromy at $\rho =0$ of the following form
\begin{equation}
\label{tildepsi}
 \widetilde \psi^j_{\pm}(\rho) \; = \;
(i\lambda_j)^{-{1\over 2}} \ \rho^{\frac{1 \pm i\lambda_j}{2}}.
\end{equation}
The  corresponding solution of the  Liouville equation reads \cite{sei}
\begin{equation}
\label{phi:0}
 \widetilde\varphi_j(\rho,\bar\rho) \; = \;
\log\left[\frac{\lambda^2_j}{\,|\rho|^2 \sin^2\left(\lambda_j\log
|\rho|\right)}\right].
\end{equation}
The metric  ${\rm e}^{\widetilde\varphi_j}d^2\rho$ has infinitely many closed geodesics:
$$
\widetilde{\cal G}_l = \{\rho \in \mathbb{C}:\lambda_j\log|\rho| =
\pi (l-{\textstyle {1\over 2}})\}\ ,
\hskip 5mm
l\in \mathbb{Z} \ ,
$$
and infinitely many singular lines:
$$
\widetilde{\cal S}_l = \{\rho \in \mathbb{C}:\lambda_j\log|\rho| = \pi l \}\ ,
\hskip 5mm
l\in \mathbb{Z} \ .
$$
Using the transformation rule (\ref{trans}) and the expansion (\ref{rho:explicit})
one can show that on $D_j \subset X$ the metric
 ${\rm e}^{\varphi}d^2z$ coincides with the pull-back of the metric
${\rm e}^{\widetilde\varphi_j}d^2\rho$ by the map $\rho_j(z)$.
As $\rho_j(D_j)$ is an open neighborhood of 0 there are infinitely many
geodesics $\widetilde{\cal G}_l $ and  singular lines
$\widetilde{\cal S}_l$ contained in $\rho_j(D_j)$.
Their inverse images  ${\cal G}_l=\rho_j^{-1}(\widetilde{\cal G}_l )$,
 ${\cal S}_l=\rho_j^{-1}(\widetilde{\cal S}_l )$, are closed singular lines
and closed geodesics of the metric ${\rm e}^{\varphi}d^2z$ on $X$.
This provides a detailed description of the singular hyperbolic geometry
in a sufficiently small neighborhood of the hyperbolic singularity:
an alternating sequence of the concentric closed geodesics and closed singular lines.
Let us stress that all these geodesic have the same length, uniquely
determined by the conformal weight:
$$
\ell_j= 2\pi\lambda_j \ .
$$

The question arises what happens to this geometry when one goes away from the singularity.
We  assume that there exists a set $\left\{\Gamma_j\right\}_{j=1}^n$ of
closed geodesics with the following properties:
\begin{itemize}
\item
$\Gamma_j$ separates $z_j$ from all other geodesics $\Gamma_i$ ($i\neq j$);
\item
the map $\rho_j$ extends to a conformal invertable map on  {\it the hole $H_j$ around } $z_j$ defined
 as the part of $\hat{\mathbb C}$
containing $z_j$ and bounded by $\Gamma_j$;
\item
the metric ${\rm e}^{\varphi}d^2z$ is regular on the surface
$
M \equiv \hat{\mathbb C} \setminus \bigcup\limits_{j=1}^n H_j $.
\end{itemize}
The assumption
is well justified by the properties of the general 3-puncture solution
and by the geometric construction of the $n$-puncture solutions \cite{HadJas4}.
In particular, it implies
that each $\Gamma_j$ is an inverse image by $\rho_j$ of one of the standard closed geodesics
${\cal G}_l$ in the $\rho$-plane. It can be parameterized as
\begin{equation}
\label{parametrization}
\gamma_j(t) \; = \; \rho_j^{-1}\left(r_j\,{\rm e}^{it}\right),
\hskip 1cm
r_j \; \equiv \; {\rm e}^{{\pi\over \lambda_j}(l + {1\over 2})},
\end{equation}
for some $l \in {\mathbb Z}$. The orientation of the  $j$-th boundary component
$\partial M_j \equiv \Gamma_j$ corresponds to the parameter  $t$ decreasing from $2\pi$ to 0.
Using (\ref{rho:explicit}) one gets for $\rho \in \rho_j(H_j)$
\begin{equation}
\label{g:explicit}
\rho_j^{-1}(\rho) \; = \; z_j + {\rm e}^{-{\vartheta_j\over \lambda_j}} \rho
- \frac{c_j}{2\Delta_j}{\rm e}^{-2{\vartheta_j\over \lambda_j}}
\rho^2 + {\cal O}(\rho^3) \ .
\end{equation}

\section{Liouville action}

The standard Liouville action on a surface $M \subset \mathbb{C}$ with
regular boundary components reads
\begin{equation}
\label{the:action}
S_{\rm\scriptscriptstyle L}\left[M,\phi\right]
\;  = \;
\frac{1}{2\pi}\int\limits_M\! d^2z\;
    \left(\partial\phi\bar\partial\phi + {\rm e}^\phi\right)
    + \frac{1}{2\pi}\int\limits_{\partial M}\!|dz|\;\kappa_z\phi,
\end{equation}
where $d^2z = \frac{i}{2}dz\wedge \bar dz$
and $\kappa_z$ is a geodesic curvature of $\partial M$ (computed in the flat metric on
the complex plane).  It yields the boundary conditions
\begin{equation}
\label{bound:conditions}
n^a\partial_a\phi + 2\kappa_z = 0
\end{equation}
and the equation of motion  (\ref{EOM}). The classical solution $\varphi(z,\bar z)$
 defines  on $M$ a hyperbolic metric ${\rm e}^\phi d^2z$ with geodesic boundaries.
If $M$ is unbounded one has to impose an appropriate asymptotic conditions
on admissible solutions. It can be done by means of a modified action
\begin{eqnarray}
\label{the:actionR}
S^\infty_{\rm\scriptscriptstyle L}\left[M,\phi\right] &=&
\lim_{R\to\infty} S^R_{\rm\scriptscriptstyle L}\left[M,\phi\right] \ ,\\
\nonumber
S^R_{\rm\scriptscriptstyle L}\left[M,\phi\right]
&=&
\frac{1}{2\pi}\int\limits_{M^R}\! d^2z\;
    \left(\partial\phi\bar\partial\phi + {\rm e}^\phi\right)
    + \frac{1}{2\pi}\int\limits_{\partial M}\!|dz|\;\kappa_z\phi +\frac{1}{\pi R}\int\limits_{|z| = R}\!|dz|\;\phi
+ 4\log R \ ,
\end{eqnarray}
where $M^R=\{z\in M:|z| \leqslant R\}$.
The presence of the additional boundary terms forces $\phi(z,\bar z)$ to
behave asymptotically as
\begin{equation}
\label{phias}
\phi(z,\bar z) \approx -2\log|z|^2
\hskip 1cm {\rm for}\hskip 1cm
|z| \to \infty \ .
\end{equation}
This implies that $T(z)$ is regular at infinity
and the limit
(\ref{the:actionR})
exists.

Let $\varphi(z,\bar z)$ denote a  solution of the Liouville equation (\ref{EOM})
 with the holomorphic component of the energy--momentum
tensor of the form (\ref{T:zz})  and satisfying the regularity conditions
formulated at the end of Sect.2.  It defines
a surface $M$ with holes $H_j$ around each hyperbolic singularity $z_j$.
The shape of $M$ depends on the conformal weights $\Delta_j$ and the location of singularities $z_j$.
The starting point of our construction is the Liouville action
$S^\infty_{\rm\scriptscriptstyle L}\left[M,\phi\,\right]$
 on this particular surface.  We shall regard $S^\infty_{\rm\scriptscriptstyle L}\left[M,\phi\,\right]$
as a functional on the space of all
conformal factors $\phi(z,\bar z)$ on $M$  with the asymptotic behavior (\ref{phias})
and satisfying the boundary conditions (\ref{bound:conditions}).
The stationary point coincides by construction with the solution
$\varphi(z,\bar z)$ restricted to $M$. The classical action
$S^\infty_{\rm\scriptscriptstyle L}\left[M,\varphi\right]$
(i.e. the action (\ref{the:actionR}) evaluated at the classical solution
$\varphi(z,\bar z)$ on $M$)\footnote{
We shall reserve the symbol $\varphi(z,\bar z)$ for the classical
solution of the equation of motion, denoting by $\phi(z,\bar z)$ a general,
``fluctuating'' field.
}
does not satisfy
the Polyakov conjecture. It turns out that the terms one has to add
to $S^\infty_{\rm\scriptscriptstyle L}\left[M,\varphi\right]$
are independent of the "fluctuating field"
$\phi(z,\bar z)$ and therefore alter neither the boundary conditions
nor the equation of motion.

On each hole $H_j$ there exists a unique flat metric with the only singularity at $z_j$
such that the boundary $\partial H_j =
\Gamma_j$ is geodesic and  its length  is  $2\pi \lambda_j$.
It can be constructed as the pull-back by $\rho_j(z)$ of the metric ${\lambda_j^2\over |\rho|^2}d^2\rho$
which yields the following formula for its  conformal factor
\begin{equation}
\label{f:definition}
\varphi_j(z,\bar z) =
\log\left[\frac{\lambda_j^2}{\left|\rho_j(z)\right|^2}
\left|\frac{d\rho_j(z)}{dz}\right|^2\right].
\end{equation}
Using the expansion (\ref{rho:explicit}) one gets
\begin{equation}
\label{asymptotic}
\varphi_j(z,\bar z) =\log \lambda_j^2  -\log|z-z_j|^2
+\frac{c_j}{2\Delta_j}(z-z_j) +
\frac{\bar{c_j}}{2\bar \Delta_j }(\bar z-\bar{z_j})
+ {\cal O}\left(|z-z_j|^2\right).
\end{equation}
Let us note that $\varphi_j(z,\bar z)$ satisfies $\mathcal{C}^1$ sewing relations
along the boundary
\begin{equation}
\label{sewing}
\varphi(z,\bar z) = \varphi_j(z,\bar z),
\hskip 5mm
n^a\partial_a \varphi(z,\bar z) = n^a\partial_a  \varphi_j(z,\bar z)\;\;\;\;{\rm for}\;\; z\in \Gamma_j \ .
\end{equation}
We define on $H_j$ the regularized classical action
$$
S^\epsilon_{\rm\scriptscriptstyle L}\left[H_j,\varphi_j\right] =
\frac{1}{2\pi} \int\limits_{H_j^\epsilon}\! d^2z\;
  \partial \varphi_j\bar\partial \varphi_j
    + \frac{1}{2\pi} \int\limits_{\Gamma_j}\!|dz|\;\kappa_z \varphi_j
+\log\epsilon \ ,
$$
where $H_j^\epsilon$ denotes $H_j$ with a disc of radii $\epsilon$ around $z_j$
cut out.
With this notation our proposal for the Liouville action in the case of hyperbolic singularities
 can be written in
the following form:
\begin{eqnarray}
\label{modified1}
S_{\rm\scriptscriptstyle L}\left[\phi\right]
&=&
\lim_{\epsilon\to 0}
S^{\epsilon}_{\rm\scriptscriptstyle L}\left[\phi\right] \ , \\
\label{modified2}
S^{\epsilon}_{\rm\scriptscriptstyle L}\left[\phi\right]
 &=&  S^\infty_{\rm\scriptscriptstyle L}\left[M,\phi\right]
+\sum\limits_{k =1}^n S^\epsilon_{\rm\scriptscriptstyle L}\left[H_k,\varphi_k\right]
- \sum\limits_{k =1}^n\lambda_k^2
\log\left|r_k^{-1}\frac{d\rho_k}{dz}(z_k)\right| \ .
\end{eqnarray}
It should be stressed that the Polyakov conjecture determines
the classical Liouville action $S_{\rm\scriptscriptstyle L}\left[\varphi\right]$
only up to an arbitrary function of
conformal weights. This freedom is tacitly assumed in the formula above.
Using the sewing relations (\ref{sewing}) one can rewrite  the classical action
$ S^{\epsilon}_{\rm\scriptscriptstyle L}\left[\varphi\right]$
(\ref{modified2})
in the form
\begin{eqnarray}
\label{classical}
 S^{\epsilon}_{\rm\scriptscriptstyle L}\left[\varphi\right]
& = & \;
{1\over 2\pi}\int\limits_M\!\!d^2z\;\left(\partial \varphi \bar\partial \varphi
+{\rm e}^\varphi \right)
+ {1\over 2\pi}\sum\limits_{k =1}^n\int\limits_{H^\epsilon_k}\!\!d^2z\;\partial \varphi_k \bar\partial \varphi_k
\\
\nonumber
&-& \sum\limits_{k =1}^n\lambda_k^2
\log\left|r_k^{-1}\frac{d\rho_k(z_k)}{dz}\right|
+ n\log\epsilon \ .
\end{eqnarray}
The modifications of the action related to the asymptotic behavior at infinity are
independent of the locations of singularities and are irrelevant for our derivation of the
Polyakov conjecture. For the sake of brevity they are suppressed  in the formula above.

\section{Polyakov conjecture}

Using the equations of motion
\begin{equation}
\label{eom:f}
\partial\bar\partial \varphi
={\textstyle \frac{1}{2}}{\rm e}^\varphi, \hskip 10mm
\partial\bar\partial \varphi_j
\;=\;0,
\end{equation}
and the sewing relations (\ref{sewing}) one gets
\begin{eqnarray}
\label{computations}
\frac{\partial}{\partial z_j} S^{\epsilon}_{\rm\scriptscriptstyle L}\left[\varphi\right]
&=&
\frac{i}{4\pi}\sum\limits_{k=1}^n \,
\int\limits_{\partial M_k}\!\!{\rm e}^\varphi
\left(\frac{\partial\gamma_k}{\partial z_j}d\bar\gamma_k
- \frac{\partial\bar\gamma_k}{\partial z_j}d\gamma_k\right)
+
\frac{i}{4\pi}\!\!\!\!\!\!
\int\limits_{\;\;\;\;|z-z_j|=\epsilon}\!\!\!\!d\bar z\;\partial \varphi_j\bar\partial \varphi_j
 \\
&+&
\frac{i}{4\pi}\sum\limits_{k=1}^n\;\!\!\!\!\!\!
\int\limits_{\;\;\;\;|z-z_k|=\epsilon}\!\!\!\frac{\partial \varphi_k}{\partial z_j}
\left(\bar\partial \varphi_k\,d\bar z - \partial \varphi_k\,dz\right)
-
\sum\limits_{k =1}^n\lambda_k^2
\frac{\partial}{\partial z_j}
\log\left|\frac{d\rho_k}{dz}(z_k)\right|.\nonumber
\end{eqnarray}
The first term on the r.h.s.\ of (\ref{computations}) results from the
change of the shape and the position of the boundary components  $\partial M_k$
 induced by the change of
$z_j$. The second one is due to the change of the position of the circle
$|z-z_j|=\epsilon$ (by construction, all the remaining ``small
holes'' preserve their positions; their radii,
equal to $\epsilon,$ are fixed). The third term follows
(after integration by parts) from the expression
\[
{1\over 2\pi}\int\limits_M\!\! d^2z\;
\left(
\partial \frac{\partial \varphi}{\partial z_j}\,\bar\partial \varphi
+
\partial \varphi\, \bar\partial\frac{\partial \varphi}{\partial z_j} +
{\rm e}^\varphi \frac{\partial \varphi}{\partial z_j}\right)
 +
{1\over 2\pi}\sum\limits_{k =1}^n
\int\limits_{H^\epsilon_k}\!\!d^2z\;\left(
\partial \frac{\partial \varphi_k}{\partial z_j}\,\bar\partial \varphi_k
+
\partial \varphi_k\, \bar\partial\frac{\partial \varphi_k}{\partial z_j}\right)
\]
resulting from the change of the integrands in (\ref{modified1}) due to
$z_j \to z_j + \delta z_j.$
Using  (\ref{parametrization}), (\ref{f:definition}), (\ref{sewing})   and
(\ref{g:explicit}) one gets:
\begin{eqnarray*}
{\rm e}^\varphi d\bar\gamma_k
\; = \;
\left.\frac{\lambda_k^2}{r_k^2}\frac{1}{(\rho^{-1}_k)'}\,d\bar \rho
\,\right|_{\bar \rho = r_k{\rm e}^{-it}}
& = &
-i\frac{\lambda_k^2}{r_k}\left({\rm e}^{{\vartheta_k\over \lambda_k}} {\rm e}^{-it} +
r_k \frac{c_k}{\Delta_k}\right)\,dt + {\cal O}\left({\rm e}^{it}\right),
\\
{\rm e}^\varphi d\gamma_k
\; =\;
\left.\frac{\lambda_k^2}{r_k^2}\frac{1}{\overline{(\rho^{-1}_k)'}}\,d\rho
\,\right|_{\rho = r_k{\rm e}^{it}}
& = &
i\frac{\lambda_k^2}{r_k}\left({\rm e}^{{\vartheta_k\over \lambda_k}} {\rm e}^{it} +
r_k \frac{\bar c_k}{\bar \Delta_k}\right)\,dt + {\cal O}\left({\rm e}^{-it}\right).
\end{eqnarray*}
From (\ref{g:explicit}) and (\ref{parametrization}) one also has:
\begin{eqnarray*}
\frac{\partial\bar\gamma_k}{\partial z_j}
& = &
\left.\frac{\partial}{\partial z_j}
\overline{\rho^{-1}_k}\,\right|_{\bar \rho = r_k{\rm e}^{-it}}
\; = \;
\frac{\partial}{\partial z_j}\left({\rm e}^{-{\vartheta_k\over \lambda_k}}\right)
r_k{\rm e}^{-it} +
{\cal O}\left({\rm e}^{-2it}\right), \\
\frac{\partial\gamma_k}{\partial z_j}
& = &
\left.\frac{\partial}{\partial z_j}
\rho^{-1}_k \,\right|_{\rho = r_k{\rm e}^{it}}
\; = \; \delta_{kj} +
\frac{\partial}{\partial z_j}\left({\rm e}^{-{\vartheta_k\over \lambda_k}}\right)
r_k{\rm e}^{it} +
{\cal O}\left({\rm e}^{2it}\right),
\end{eqnarray*}
and
\begin{eqnarray}
\label{comp1}
\frac{i}{4\pi}\sum\limits_{k=1}^n
\int\limits_{\partial M_k}\!\!{\rm e}^\varphi
\left(\frac{\partial\gamma_k}{\partial z_j}d\bar\gamma_k
- \frac{\partial\bar\gamma_k}{\partial z_j}d\gamma_k\right)
& = &
-\sum\limits_{k=1}^n {\lambda_k^2\over 2\pi}
\int\limits_{0}^{2\pi}\!\!dt
\left[ \frac{c_k}{2\Delta_k}\delta_{kj} +
{\rm e}^{{\vartheta_k\over \lambda_k}}
\frac{\partial}{\partial z_j}\left({\rm e}^{-{\vartheta_k\over \lambda_k}}\right)
 + \ldots \right] \nonumber \\
& = & -\frac{\lambda_j^2}{2\Delta_j}c_j +
\sum\limits_{k=1}^n\lambda_k\frac{\partial}{\partial z_j} \vartheta_k \ .
\end{eqnarray}
The terms in the first line of (\ref{comp1}) denoted by dots contain
non-zero, integer powers of ${\rm e}^{it}$ and vanish upon
integration.
The expansion (\ref{asymptotic}) implies
\begin{eqnarray*}
\partial \varphi_k &=& -\frac{1}{z-z_k} + \frac{c_k}{2\Delta_k} + {\cal
O}(z-z_k)\ ,\\
\bar \partial \varphi_k &=& -\frac{1}{\bar z-\bar z_k} +
\frac{\bar c_k}{2\bar \Delta_k} + {\cal O}(\bar z-\bar z_k)\ ,
\\
\frac{\partial}{\partial z_j}\varphi_k &=&
\delta_{kj}\left(\frac{1}{z-z_k} - \frac{c_k}{2\Delta_k}\right) +
{\cal O}(|z-z_k|)\ .
\end{eqnarray*}
Hence, up to the terms that vanish in the limit $\epsilon \to 0$ :
\begin{equation}
\label{comp2}
\frac{i}{4\pi}\!\!\!\!\!\!
\int\limits_{\;\;\;\;|z-z_j|=\epsilon}\!\!\!\!\!\!d\bar z\;\partial \varphi_j\bar\partial \varphi_j
 \; +\;
\frac{i}{4\pi}\sum\limits_{k=1}^n\!\!\!
\int\limits_{\;\;\;\;|z-z_k|=\epsilon}\!\!\!\!\!\!\frac{\partial \varphi_k}{\partial z_j}
\left(\bar\partial \varphi_k\,d\bar z - \partial \varphi_k\,dz\right)
\; = \;
-\frac{c_j}{2\Delta_j}\ .
\end{equation}
Substituting (\ref{comp1}) and (\ref{comp2}) in (\ref{computations}) and taking into account
the relation
$$
\log\left|\frac{d\rho_k}{dz}(z_k)\right|= {\vartheta_k\over \lambda_k}
$$
one finally gets
\begin{equation}
\label{Polyakov}
\frac{\partial}{\partial z_j} S_{\rm\scriptscriptstyle L}\left[\varphi\right]
\; = \; \lim_{\epsilon\to 0}
\frac{\partial}{\partial z_j} S^{\epsilon}_{\rm\scriptscriptstyle L}\left[\varphi\right]
\; = \; - c_j \ .
\end{equation}
\section*{Acknowledgements}
The work of L.H. was supported by the EC IHP network
HPRN-CT-1999-000161.
\noindent
Laboratoire de Physique Th{\'e}orique is Unit{\'e} Mixte du CNRS
UMR 8627.

\end{document}